\documentclass[conference,a4paper]{IEEEtran}
\usepackage[utf8]{inputenc} 
\usepackage[T1]{fontenc}
\usepackage{url}
\usepackage{ifthen}
\usepackage{cite}
\usepackage[cmex10]{amsmath} 
\usepackage{amssymb}
\usepackage{graphicx}
\usepackage{multirow}
\usepackage{array}
\usepackage{algorithm}
\usepackage{subfig}
\usepackage{makecell}
\usepackage{algpseudocode}
\usepackage{comment}
\usepackage{caption}
\usepackage[english]{babel}
\usepackage{romannum}
\usepackage{etoolbox}
\makeatletter
\patchcmd{\@makecaption}
  {\scshape}
  {}
  {}
  {}
\makeatother

\newtheorem{thm}{Theorem}

\newtheorem{exmp}{Example}

\newtheorem{rem}{Remark}

\begin{document}
	\title{ Maddah-Ali-Niesen Scheme for  Multi-access Coded Caching} 
	\author{%
		\IEEEauthorblockN{Pooja Nayak Muralidhar, Digvijay Katyal and B. Sundar Rajan \\
Department of Electrical Communication Engineering, Indian Institute of Science, Bengaluru 560012, KA, India \\
			E-mail: \{poojam,digvijayk,bsrajan\}@iisc.ac.in}
	}
	\maketitle
\begin{abstract}
The well known Maddah-Ali-Niesen (MAN) coded caching scheme for users with dedicated cache is extended for use in multi-access coded cache scheme where the number of users need not be same as the number of caches in the system. The well known MAN scheme is recoverable as a special case of the multi-access system considered.  The performance of this scheme is compared with the existing works on multi-access coded caching. To be able to compare the performance of different multi-access schemes with different number of users for the same number of caches, the terminology of per user rate (rate divided by the number of users)  introduced in \cite{KNS} is used.
\end{abstract}
\section{Introduction}
\label{sec1}
	Due to the increasing number of multimedia applications like video on demand, tremendous growth in the consumption of data has been observed in recent years. The seminal work of \cite{MaN} showed that jointly designing content placement and delivery, also known as coded caching significantly improves content delivery rate requirements. However in reality, due to practical constraints the subpacketization levels in \cite{MaN} is not feasible due to its exponentially increasing nature,  with respect to the increasing number of users. The quest for coded caching schemes with practical subpacketization levels started, and a wide variety of schemes using different constructions of placement delivery arrays\cite{YCTC}, line graphs of bipartite graphs\cite{PK}, linear block codes\cite{TaR}, block designs \cite{ASK} etc, were found. 
	
	Parallelly, this technique of jointly designing placement and delivery were explored in other type of networks like device to device networks (D2D) \cite{WCYT}, Combination networks\cite{YWY}, networks with shared cache \cite{AL}, \cite{SBP2}, \cite{AAA} etc.
	The case of networks with shared caches is particularly interesting since in these types of networks users can share the caches and hence in a practical perspective it helps in efficient utilization of memory. Coded caching in networks where demands are non uniform i.e., files with different popularities \cite{NMa}, \cite{HKD} has also been well studied. However we will be limiting ourselves to the cases where all the files are equipopular and demands are distinct.

	An interesting and practical type of network scenario is a multi-access network where multiple users can access the same cache and multiple caches can be accessed by the same user \cite{HKD},\cite{RaK3},\cite{SBP}, \cite{KNS}, \cite{CLWZC}, \cite{SRIC}, \cite{SR}. The scenario considered in \cite{HKD},\cite{RaK3}, \cite{SBP}, \cite{CLWZC}, \cite{SRIC} and \cite{SR} has $K$ users and $K$ caches and a “sliding window” approach, where user $k$ accesses caches $k,\;k + 1, \dots ,\;k + r - 1$ for some $r \in {1, 2, . . . , K}$, using a cyclic wraparound to preserve symmetry is used. Here, $r$ is called cache access degree i.e. the number of caches a user has access to. 

In the following subsection we provide a brief survey of the schemes that are known in the literature of multi-access networks.
\subsection*{Known schemes in multi-access}

\subsubsection{Hachem-Karamchandani-Diggavi (HKD) Scheme \cite{HKD}}
    Multi-access setups were introduced in the work of \cite{HKD} where caching and delivery in a decentralized setting with multi-access and multi-level popularities under consideration.The authors give rate memory trade offs for a multi-level access model (content is divided into discrete levels based on popularity and users are required to connect to a certain number of access points based on the popularity of the file they have requested) with multi user setups (user can access multiple caches). In the case of a multi-user, multi-access model with a single-level caching system, with $N$ files, $K$ caches and $K$ users grouped into  $U$ users in a group with access degree $r$, such that $N \geq K U$ and $r$ divides $K$, and a cache memory of $M \in [0, \frac{N}{r}]$, in the decentralized assumption, considered in the paper, the achievable rate is given by
    $$R =U.\min\Bigl\{\frac{N}{M},K\Bigr\}\Bigl(1 - \frac{r M}{N}\Bigr).$$
Also when $r$ does not divide $K$, four times the above expression can be achieved and \\
    $$R = 0\;\; \text{if}\; M > \frac{N}{r}.$$ 
In \cite{HKD}, the scheme proposed is decentralized. However the scheme can be extended to get a centralized scheme (also indicated in \cite{RaK3}) with the rate achieved given by
\begin{equation}
\label{RateHKD}
    R = \frac{K - \frac{KrM}{N}}{1 + \frac{KM}{N}}.
\end{equation}  
Henceforth  we refer to this scheme as the HKD scheme.
\subsubsection{Reddy-Karamchandani (RK) Scheme \cite{RaK3} }
    The scheme proposed in \cite{RaK3} supports a multi-access setup with $K$ users and $K$ caches with each user connected to $r$ consecutive caches in a cyclic manner. The rate for this scheme for $M = i\frac{N}{K}$ where $i \in {0 \cup [\lceil \frac{K}{r} \rceil]}$ is given by the expression\\
     \begin{align*}
        R &=  K \Big{(1-r\;\frac{M}{N}\Big)}^2 \text{ for } i \in {0 \cup \Big[\Bigl\lfloor \frac{K}{r} \Bigr\rfloor\Big]} 
       \text{ and }\\
        R &= 0 \text{ for } i = \Bigl \lceil \frac{K}{r} \Bigr \rceil
    \end{align*}
    
    A generic lower bound on the optimal rate for any such multi-access setup with $r \geq \frac{K}{2}$, under the restriction of uncoded placement is derived as
    \begin{multline*}
 R_{lb}(M)=\\\begin{cases}
     K - \Big[ K - \frac{(K-r)(K-r+1)}{2K}\Big]\frac{MK}{N} & \text{, if $0 \leq M \leq \frac{N}{K}$}\\
    \frac{(K-r)(K-r+1)}{2K}\Big(2 - \frac{MK}{N}\Big) & \text{, if $\frac{N}{K} \leq M \leq \frac{2N}{K}$}\\
    0 & \text{, if $M \geq \frac{2N}{K}$}.
  \end{cases}
\end{multline*}
    The rate achieved by this scheme is proved to be order optimal with the multiplicative gap between the achievable rate and lower bound, at most $2$ for $r \geq K/2.$
    The scheme is also found to be optimal for the special cases, $r = K-1$, $r = K-2$, $r= K-3$ with $K$ even and $r = K-\frac{K}{s}+1$ for some positive integer $s$. Hereafter, we refer to this scheme as the RK scheme.
\subsubsection{Serbetci-Parinello-Elia (SPE) Scheme\cite{SBP}}
    The work in \cite{SBP} also deals with the same problem setup as in the previous cases and provides two new schemes which can serve, on average, more than $K\gamma + 1$ users at a time and for the special case of $r = \frac{K-1}{K\gamma}$, the achieved gain is proved to be optimal under uncoded cache placement where $\gamma = \frac{M}{N},\;\gamma \in \{\frac{1}{K}, \frac{2}{K},\dots,1\}$. The general scheme is proposed for the case $K\gamma = 2$. The subpacketization for this scheme is given by $F =\frac{K(K - 2r + 2)}{4}$ and it can be noted that the numerator should be divisible by $4$ and $r<\frac{(K+2)}{2}$ for the number of subpackets to be a positive integer greater than 0. We refer to this scheme as the SPE scheme.    
 \subsubsection{Scheme using Cross Resolvable Designs (CRD) \cite{KNS}}
    In \cite{KNS} the authors develop a multi-access coded caching scheme from a specific type of resolvable designs called cross resolvable designs. The number of users supported in this scheme is  higher than the other existing schemes for the same number of caches and for practically realizable subpacketization levels. To compare the performance with other schemes, the authors introduce the notion of per user rate or rate per user obtained by normalizing the rate $R$ with the number of users $K$ supported, i.e., rate per user is $\frac{R}{K}.$ We refer to this scheme as the CRD scheme.
\subsubsection{Cheng-Liang-Wan-Zhang-Caire (CLWZC) Scheme \cite{CLWZC}}
	The work of \cite{CLWZC} proposes a novel transformation approach to extend the Maddah Ali Niesen scheme to the multiaccess caching systems, such that the load expression of the scheme by \cite{HKD} remains achievable even when $r$ does not divide $K$. This work considers only the multi-access setup with cyclic wraparound where  each user has access to $r$ neighboring cache-nodes with a cyclic wraparound setup.  The rate expression for this scheme is same as that for the centralized HKD scheme. The subpacketization required is $K\binom{K - t(r-1)}{t}$, where $t = \frac{KM}{N}$.
\subsection{Sasi Rajan Scheme 1\cite{SRIC}}
In \cite{SRIC}, multi-access schemes based on cyclic setup with some improvement over some of the cases in RK scheme and CLWZC scheme has been proposed. The scheme proposed exists for the cases when normalized capacity  $\gamma$, where $\gamma\;\in\;\{\frac{t}{K}\;:\;\text{gcd}(t, K) = 1, t \in [1, K]\}$. The rates achieved here, are given below.
\begin{itemize}
	\item  $\mathcal{R}_{SR1}(\gamma) = \frac{1}{K}$,  if  $(K- t r)=1.$ 
	\item If $(K- t r)$ is even, then 
	$\mathcal{R}_{SR1}(\gamma) = 2 \sum_{i=\frac{K- t r}{2}+1}^{K- t r}\frac{1}{1+\left \lceil \frac{t r}{i} \right \rceil}.$
	\item If $(K- tr)>1$ is odd, then 
	{\small
		$\mathcal{R}_{SR1}(\gamma) = \frac{1}{ \left (\left \lceil  \frac{2k L}{K-t r+1} \right \rceil+1 \right )}+\sum_{i=\frac{K- t r+3}{2}}^{K- t r}\frac{2}{1+\left \lceil \frac{t r}{i} \right \rceil}. $}
\end{itemize}

The subpacketization achieved here is atmost $K^2$. 
The scheme outperforms the RK scheme in all cases other than the ones achieving optimality. The scheme proposed performs better than the CLWZC scheme when $r \geq \Big(1 - \frac{1}{t}\Big)\frac{K}{t}$. 
This scheme is a generalisation of one of the cases considered in \cite{SBP}, i.e., when $r = \frac{K-1}{K\gamma}$.We will refer to this scheme as the SR scheme 1. 
\subsection{Sasi Rajan Scheme 2\cite{SR}}
In \cite{SR}, multi-access schemes based on placement delivery arrays have been proposed in the work of \cite{SR}, where sub-packetization level varies only linearly with the number of users. The rate achieved by this scheme is
$$ R_{SR2} = \frac{(K - tr)(K- tr + t)}{2K}$$
where 
$\gamma\;\in\;\{\frac{t}{K}\;: t|K,(K - tr + t)| K, t \in [1, K]\}$ and
and subpacketization $F = K$. The scheme of \cite{SR} is found to be better than the CLWZC scheme for $r \geq \frac{K(t-1)}{t(t+1)} + 1$. This scheme  also has performance better than or equal to that of RK scheme. We will refer to this scheme as the SR scheme 2. 
\subsection{Our Contributions}
	In this work a coded caching scheme for multi-access networks is proposed with the number of users much larger than the number of caches. The proposed scheme is extension of the MAN scheme and the MAN scheme is can be obtained as a special case. The performance of this scheme is compared with the existing works. To be able to compare the performance of different multi-access schemes with different number of users for the same number of caches, the terminology of per user rate $\frac{R}{K}$ introduced in \cite{KNS} is used.
   
{\it Notations:} The set $\{1,2,...n\}$ is denoted as $[n]$ $|\mathcal{X}|$ denotes cardinality of the set $\mathcal{X}$.

\section{The Proposed Scheme}
\label{sec2}
\begin{figure}
        \begin{center}
                \includegraphics[width=7cm,height=6cm]{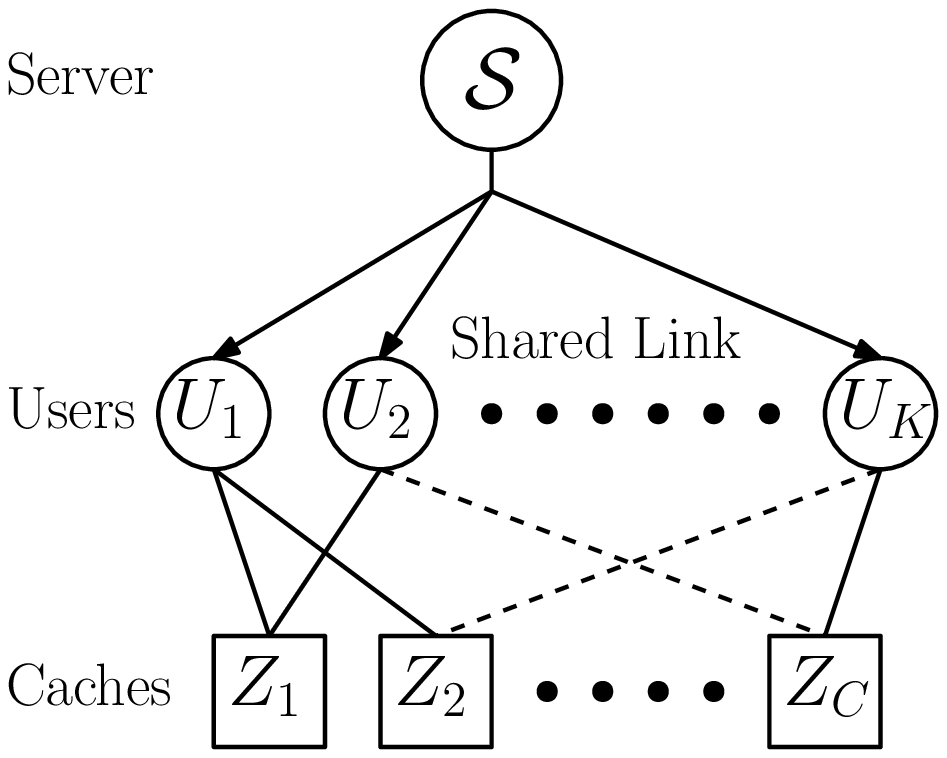}
                \caption {Problem setup}
                \label{setup}
        \end{center}
\end{figure}

        Our problem set up is as as shown in Fig. \ref{setup}. Let $\mathcal{K}$ denote the set of $K$ users in the network  connected via an error free shared link to a server $\mathcal{S}$ storing $N$ files $(N \geq K)$ denoted as $W_1, W_2, W_3,\dots,W_N$ each of  unit size. Each user can access a unique set of $r$ caches (cache access degree) out of $C$ caches, each capable of storing $M$ files. $\mathcal{Z}_k$ denotes the content in cache $k$, and we assume that each user has an unlimited capacity link to the caches that it is connected to.
Let $\mathcal{C}$ denote the set of $C$ caches. Since a user has access to a unique subset of $r$ caches, every user can be uniquely represented with a $r$ sized subset of $\mathcal{C}$. From now on, we denote the user by the $r$ sized subset of caches it is connected to.  Let $d_{\mathcal{U}}$, where  $\mathcal{U} \subset \mathcal{C};\;|\mathcal{U}| = r$ denote the demand of the user connected to the set of caches represented by $\mathcal{U}$. Hence, the maximum number of users $K$ in this scheme is $C\choose r$. Our scheme works for any number of users as long as a user is associated with a unique subset of $r$ caches, i.e., no two users are accessing the same set of $r$ caches. For concreteness we assume that $K= {C \choose r}.$  The scheme works in two phases described below:

\subsubsection*{Placement Phase} 
Let $t = \frac{CM}{N}$ be an integer. The server divides each file $W_{i}$ into $C \choose t$ subfiles and in the $k^{th}$ cache places the content  given by
	$$\mathcal{Z}_k = \{W_{i,\mathcal{T}}:k \in \mathcal{T}, \; \mathcal{T} \subset \mathcal{C},\;|\mathcal{T}| = t,\;\; \forall \;i \in [N]\}.$$ 
	It is seen  that after the placement, the size of the content stored in the cache is equal to $\frac{{{C-1} \choose {t-1}}}{{{C} \choose {t}}} N$ = $\frac{t N}{C}$ = $M$.
	
	Note that a user may find the same subfiles of a file in more than one cache it has access to. Let $M'$ denote the size of the files that a user has access to. We have
	
\begin{multline*}
\frac{M'}{N}= \sum_{i=1}^r \frac{|\mathcal{Z}_i|}{\binom{C}{t}}-\sum_{1\leq i_1<i_2\leq r}^r \frac{|\mathcal{Z}_{i_1}\cap\mathcal{Z}_{i_2}|}{\binom{C}{t}}+\dots +\\(-1)^{s+1}\sum_{1\leq i_1<\dots<i_s\leq r}^{r} \frac{|\mathcal{Z}_{i_1}\cap\dots \cap\mathcal{Z}_{i_s}|}{\binom{C}{t}}
+\dots+\\ (-1)^{r+1}\frac{|\mathcal{Z}_{1}\cap\dots\cap\mathcal{Z}_{r}|}{\binom{C}{t}}
\end{multline*}
\noindent where $\mathcal{Z}_i,\;i\in[r]$ are distinct $r$ caches. The above expression simplifies to 
\begin{multline*}
    \frac{M'}{N}= \dfrac{\binom{r}{1}\binom{C-1}{t-1}}{\binom{C}{t}}-\dfrac{\binom{r}{2}\binom{C-2}{t-2}}{\binom{C}{t}}+\dots+(-1)^{(s+1)}\dfrac{\binom{r}{s}\binom{C-s}{t-s}}{\binom{C}{t}}
\\+\dots+(-1)^{(r+1)}\dfrac{\binom{r}{r}\binom{C-r}{t-r}}{\binom{C}{t}}
\end{multline*}
$$\frac{M'}{N} = \frac{1}{\binom{C}{t}}\left(\sum_{n=1}^{r} (-1)^{n+1}\binom{r}{n}\binom{C-n}{t-n}\right)$$

\begin{rem}
	Note that the placement is same as the placement of Maddah Ali Niesen scheme in shared link networks \cite{MaN} with the difference that in \cite{MaN}, each user is equipped with a dedicated cache.
\end{rem}

\subsubsection*{Delivery Phase}
	For each such subset $S \subset \mathcal{C}$ of cardinality $|S| = t + r$, the server transmits $$	\mathcal{Y}_{S} = \bigoplus_{\underset{{|\mathcal{U}|\;=\;r}}{\mathcal{U}\;\subset\;S}} W_{d_\mathcal{U},\;S\setminus\mathcal{U}}$$

Now we prove that with the above delivery scheme every user will be able to decode the file it wants to. 
\begin{IEEEproof}
First, we note that when $\mathcal{U} \cap \mathcal{T} = \emptyset$, where $\mathcal{T} \subset \mathcal{C},\;|\mathcal{T}| = t$, then the user $\mathcal{U}$ does not have access to any cache $c$ where, $c \in \mathcal{T}$ and $c \notin \mathcal{U}$. A user $\mathcal{U}$ has access to a subfile indexed by $\mathcal{T}$ iff $\mathcal{U} \cap \mathcal{T} \neq \emptyset$. This is because, when $\mathcal{U} \cap \mathcal{T} \neq \emptyset$, there exists a cache $c \in \mathcal{U} \cap \mathcal{T}$, through which $\mathcal{U}$ can access the subfile indexed by $\mathcal{T}$.

Consider the delivery algorithm mentioned  above. We now argue that each user can successfully recover its requested message. Consider the transmission by the server
   $$\bigoplus_{\underset{{|\mathcal{U}|\;=\; r}}{\mathcal{U}\;\subset\;S}} W_{d_\mathcal{U},\;S\setminus\mathcal{U}}$$
   corresponding to the subset of caches, $S \subset \mathcal{C}$ with $|S| = t + r$. Consider a user $\mathcal{U} \subset S ;\; |\mathcal{U}| = r$. The user $\mathcal{U}$ already has access to the subfiles $W_{d_{\mathcal{U}},\;S\setminus\mathcal{V}}$  for any other user indexed by $\mathcal{V} \subset \{ \mathcal{S} : \mathcal{V} \neq  \mathcal{U}\} ;\; |\mathcal{V}| = r$ since, $\{S\setminus\mathcal{V}\} \cap \mathcal{U} \neq \emptyset$.
   Hence it can retrieve the subfile $W_{d_{\mathcal{U}}, ~ S\setminus\{\mathcal{U}\}}$ from the transmission corresponding to $S$. Likewise, from all such transmissions corresponding to $S$, with $\mathcal{U} \subset S$, user $\mathcal{U}$ gets the missing subfiles of $W_{d_{\mathcal{U}}}$ 
   sent over the shared link. Since this is true for every such subset $S$, any user $\mathcal{U} \subset \mathcal{C} ;\; |\mathcal{U}| = r$  can recover all missing subfiles.\\
	\end{IEEEproof}

\begin{thm}
For every integer $t = \frac{CM}{N},$ 	the above placement and delivery results in a multi-access scheme with $C$ caches, access degree $r$, $C\choose r$ users, subpacketization $F = {C\choose t}$, coding gain $g = {{t+r} \choose {r}}$, and rate $R = \frac{{ C\choose {t+r}}}{{ C\choose t}}.$
\end{thm}

\begin{IEEEproof}
	According to the proposed scheme, the size of each subpacket is $\frac{1}{{{C}\choose {t}}}$. The total number of transmissions is the total number of choices of $S = {C \choose {t+r}}$. The coding gain, defined as the total number of users benefited from each transmission is thus, the number of $r$ sized subsets of $\mathcal{C}$ (number of users) we can choose from a particular  $S$. Thus rate $R = \frac{{ C\choose {t+r}}}{{ C\choose t}}$ and the coding gain in this scheme is ${{t+r} \choose {r}}$
\end{IEEEproof}

\begin{rem}
	The proposed scheme can be designed for any integer $t = \frac{CM}{N}$, and the rate points in between can be achieved through memory sharing. So the proposed scheme exists for any multi-access network with $N$ files, $C$ caches equipped with memories of size $M$ file units each and access degree $r$.
\end{rem}

\begin{rem}
	Note that in this setup, since every user is associated with a unique subset of caches, even if some users leave the network, the transmissions involving the existing users continue to hold. The only constraint is that, any user in the network should be uniquely associated with an $r$ subset of caches. So our setup is dynamic in the sense that users can join and leave the system whenever they want, provided the association with the caches is unique. $C \choose r$ is simply the maximum number of users associated with distinct $r$ subsets of caches that can be supported by the described the placement and delivery scheme. 
\end{rem}
	
\begin{rem}
	In all the multi-access schemes in literature except the one derived from cross resolvable designs, the $r$ caches connected to any user, store disjoint contents. As a consequence, it is seen that in schemes with cyclic setups, the rate $R$ becomes $0$ when $\frac{M}{N} \geq \frac{1}{r}$, since the user gets all the $N$ files from the $r$ caches connected to it. However, in these schemes, it is possible to enforce the constraint of making $r$ caches store disjoint contents since the number of users in the network is only $C$. In the setup considered here, since the number of users is more than $C$ , the same subfiles can be stored in multiple caches connected to a user and hence rate does not become $0$, when $\frac{M}{N} \geq \frac{1}{r}$. Though this redundancy does appear like a waste of memory, since the number of users supported is large in the network, multi-casting opportunities increase resulting in low per user rates.
\end{rem}
	
\begin{rem}
    In the proposed scheme, the contents placed in the caches connected to any user in the setup are disjoint when $t=1$, since the contents placed in the caches itself is disjoint in this special case.
    $$Z_i \cap Z_j = \emptyset,\; \; \forall\; i,j \in \mathcal{C}$$
\end{rem}

	It can be noted that $t=1$, is the only case when contents connected to a user become disjoint, like in the existing schemes in literature where caches connected to a user always hold disjoint contents with the exception of the CRD coded caching scheme.
	
\subsection{Examples}
In the examples below, we assume that the users are numbered by  lexicographically ordered $r$ subsets. For example if $r = 2$,  $C = 4$, then  User 1 corresponds to the subset $\{1, 2\}$, User 2 to $\{1, 3\}$, User 3 to $\{1, 4\}$, User 4 to $\{2, 3\}$, User 5 to $\{2, 4\}$, and User 6 to $\{3, 4\}$. The request vector $(d_1,d_2,\dots,d_K) \in N^K$ denotes the tuple containing demands of the users $1,2,..,K$. The following three examples correspond to the cases $t=1,$ $t \neq r,$ and $t=r$ respectively.
\begin{exmp}
\label{exmp1}
Consider a multi-access setup with $C = 4$ caches, and cache access degree $r = 2$. We allow all possible combinations of access to $r$ caches. So number of users $K = {C \choose r} = {4 \choose 2} = 6.$ For $t=1$ the subpacketization is $F = {C \choose t} = {4 \choose 1} = 4.$ The subfiles are
	$W_{i,1}$, $W_{i,2}$, $W_{i,3}$, $W_{i,4},\; \forall \;  i \in [N].$ The cache placement is:
    	\begin{align*}
		\mathcal{Z}_1 =\;&\; \{W_{1,1},W_{2,1},W_{3,1},W_{4,1},W_{5,1},W_{6,1}\}\\
		\mathcal{Z}_2 =\;&\; \{W_{1,2},W_{2,2},W_{3,2},W_{4,2},W_{5,2},W_{6,2}\}\\
		\mathcal{Z}_3 =\;&\; \{W_{1,3},W_{2,3},W_{3,3},W_{4,3},W_{5,3},W_{6,3}\}\\
		\mathcal{Z}_4 =\;&\; \{W_{1,4},W_{2,4},W_{3,4},W_{4,4},W_{5,4},W_{6,4}\}\\
	    \end{align*}
	Let the request vector be $(1,2,3,4,5,6)$. The transmissions are:
		\begin{align*}
		\mathcal{Y}_{\{1,2,3\}} &= W_{1,3}\oplus W_{2,2}\oplus W_{4,1}\\
		\mathcal{Y}_{\{1,2,4\}} &= W_{1,4}\oplus W_{5,1}\oplus W_{3,2}\\
		\mathcal{Y}_{\{1,3,4\}} &= W_{2,4}\oplus W_{3,3}\oplus W_{6,1}\\
		\mathcal{Y}_{\{2,3,4\}} &= W_{5,3}\oplus W_{6,2}\oplus W_{4,4}\\
		\end{align*}

\end{exmp}
\begin{exmp}
\label{exmp2}
Consider a multi-access setup with $C = 5$ caches, and cache access degree $r = 3$, $t = 2$. Number of users $K$ = $C \choose r$ = $5 \choose 3$ = $10$ and subpacketization is $F$ = $C \choose t$ = $5 \choose 2$ = $10$. The subfiles are
	$W_{i,\{1,2\}}$, $W_{i,\{1,3\}}$, $W_{i,\{1,4\}}$, $W_{i,\{1,5\}}$, $W_{i,\{2,3\}}$, $W_{i,\{2,4\}}$, 
	$W_{i,\{2,5\}}$, $W_{i,\{3,4\}}$, $W_{i,\{3,5\}}$, $W_{i,\{4,5\}}, \; \forall \;  i \in [N].$ The cache placement is:
    	\begin{align*}
		\mathcal{Z}_1 =\;&\;\{W_{i,\{1,2\}},W_{i,\{1,3\}},W_{i,\{1,4\}},W_{i,\{1,5\}},\;\; \forall \;  i \in [N]\}\\
		\mathcal{Z}_2 =\;&\;\{W_{i,\{1,2\}},W_{i,\{2,3\}},W_{i,\{2,4\}},W_{i,\{2,5\}},\;\; \forall \;  i \in [N]\} \\
		\mathcal{Z}_3 =\;&\;\{W_{i,\{1,3\}},W_{i,\{2,3\}},W_{i,\{3,4\}},W_{i,\{3,5\}},\;\; \forall \;  i \in [N]\} \\
		\mathcal{Z}_4 =\;&\;\{W_{i,\{1,4\}},W_{i,\{2,4\}},W_{i,\{3,4\}},W_{i,\{4,5\}},\;\; \forall \;  i \in [N]\} \\
	    \mathcal{Z}_5 =\;&\;\{W_{i,\{1,5\}},W_{i,\{2,5\}},W_{i,\{3,5\}},W_{i,\{4,5\}},\;\; \forall \;  i \in [N]\} 
	    \end{align*}
	Let the request vector be $(1,2,3,4,5,6,7,8,9,10)$. The transmissions are:
		\begin{multline*}
		    \mathcal{Y}_{\{1,2,3,4,5\}} = W_{1,\{4,5\}}\oplus W_{2,\{3,5\}}\oplus W_{3,\{3,4\}} \oplus\\
		 W_{4,\{2,5\}}\oplus
		W_{5,\{2,4\}}\oplus W_{6,\{2,3\}}  \oplus
	    W_{7,\{1,5\}}\oplus\\ W_{8,\{1,4\}}\oplus
	    W_{9,\{1,3\}} \oplus W_{10,\{1,2\}}
		\end{multline*}
	\end{exmp}
\begin{figure}[H]
	\centering
	\subfloat[$\frac{M}{N}  < \frac{1}{r}$]{
	\includegraphics[clip,width=0.8\columnwidth]{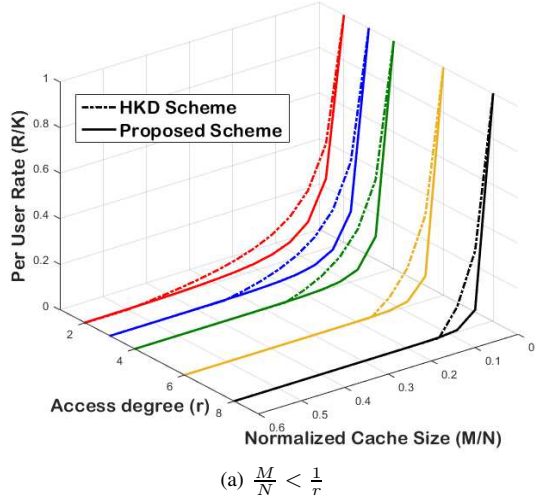}
	\label{HKDcomp_RK1}}\\
	\subfloat[$\frac{M}{N}  \geq \frac{1}{r}$]{
	\includegraphics[clip,width=0.8\columnwidth]{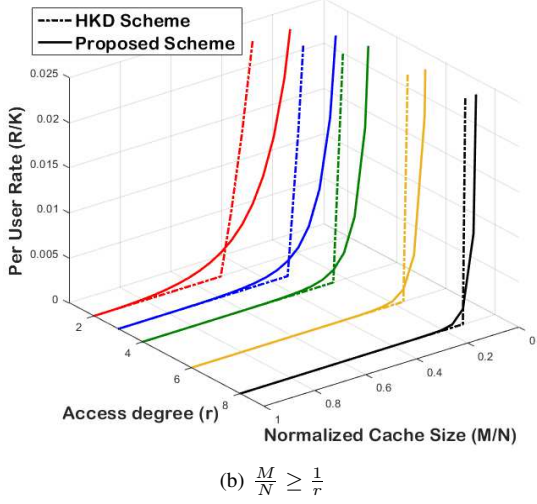}
	\label{HKDcomp_RK2}}
\caption {Comparison with the HKD scheme for the number of caches, $C = 24$. The two schemes are compared for different values of $r$ i.e $r\in\{2,3,4,6,8\}$.}
\end{figure}
\begin{exmp}
\label{exmp3}
		Consider a multi-access setup with $C = 5$ caches , and cache access degree $r = 2$, $t = 2$. Number of users $K$ = $C \choose r$ = $5 \choose 2$ = $10$ and subpacketization is $F$ = $C \choose t$ = $5 \choose 2$ = $10$. The subfiles are
	$W_{i,\{1,2\}}$, $W_{i,\{1,3\}}$, $W_{i,\{1,4\}}$, $W_{i,\{1,5\}}$, $W_{i,\{2,3\}}$, $W_{i,\{2,4\}}$, 
	$W_{i,\{2,5\}}$, $W_{i,\{3,4\}}$, $W_{i,\{3,5\}}$, $W_{i,\{4,5\}}, \; \forall \;  i \in [N].$ The cache placement is:
    	\begin{align*}
		\mathcal{Z}_1 =\;&\;\{W_{i,\{1,2\}},W_{i,\{1,3\}},W_{i,\{1,4\}},W_{i,\{1,5\}},\;\; \forall \;  i \in [N]\}\\
		\mathcal{Z}_2 =\;&\;\{W_{i,\{1,2\}},W_{i,\{2,3\}},W_{i,\{2,4\}},W_{i,\{2,5\}},\;\; \forall \;  i \in [N] \}\\
		\mathcal{Z}_3 =\;&\;\{W_{i,\{1,3\}},W_{i,\{2,3\}},W_{i,\{3,4\}},W_{i,\{3,5\}},\;\;  \forall \;  i \in [N]\}\\
		\mathcal{Z}_4 =\;&\;\{W_{i,\{1,4\}},W_{i,\{2,4\}},W_{i,\{3,4\}},W_{i,\{4,5\}},\;\;  \forall \;  i \in [N]\}\\
	    \mathcal{Z}_5 =\;&\;\{W_{i,\{1,5\}},W_{i,\{2,5\}},W_{i,\{3,5\}},W_{i,\{4,5\}},\;\;  \forall \;  i \in [N]\}\\
	    \end{align*}
\begin{figure}[H]
	\centering
	\subfloat[In the plot shown above, the two schemes are compared for different values of $t = \frac{CM}{N}$ i.e $t\in\{2,3,4,5\}$, $r = 2$.]{
	\includegraphics[clip,width=0.8\columnwidth]{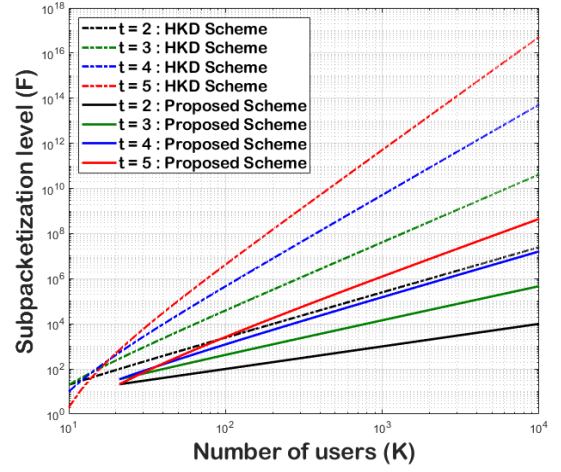}
	\label{HKDcomp_rconst}}\\
	\subfloat[In the plot shown above, the two schemes are compared for different values of $r$ i.e $r\in\{2,3,4,5\}$, $t = \frac{CM}{N} = 2$.]{
	\includegraphics[clip,width=0.8\columnwidth]{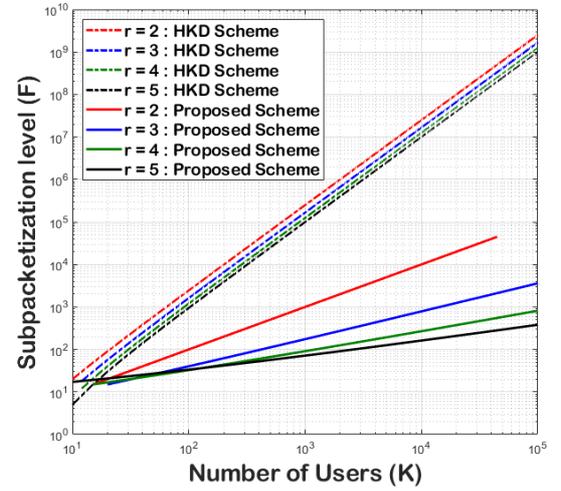}
	\label{HKDcomp_tconst}}
\caption {Comparison with the HKD scheme in terms of subpacketization.}
\end{figure}

    Let the request vector be $(1,2,3,4,5,6,7,8,9,10)$. The transmissions are:
		\begin{multline*}
		    \mathcal{Y}_{\{1,2,3,4\}} = W_{1,\{3,4\}}\oplus W_{2,\{2,4\}}\oplus W_{3,\{2,3\}}\oplus W_{5,\{1,4\}}\oplus\\ W_{6,\{1,3\}}\oplus W_{8,\{1,2\}}
		\end{multline*}
		\begin{multline*}
		    \mathcal{Y}_{\{1,2,3,5\}} = W_{1,\{3,5\}}\oplus W_{2,\{2,5\}}\oplus W_{4,\{2,3\}}\oplus W_{5,\{1,5\}}\oplus\\ W_{7,\{1,3\}}\oplus W_{9,\{1,2\}}
		\end{multline*}
		\begin{multline*}
		    \mathcal{Y}_{\{1,2,4,5\}} = W_{1,\{4,5\}}\oplus W_{3,\{2,5\}}\oplus W_{4,\{2,4\}}\oplus W_{6,\{1,5\}}\oplus\\ W_{7,\{1,4\}}\oplus W_{10,\{1,2\}}
		\end{multline*}
		\begin{multline*}
		    \mathcal{Y}_{\{1,3,4,5\}} = W_{2,\{4,5\}}\oplus W_{3,\{3,5\}}\oplus W_{4,\{3,4\}}\oplus W_{8,\{1,5\}}\oplus\\ W_{9,\{1,4\}}\oplus W_{10,\{1,3\}}
		\end{multline*}
		\begin{multline*}
		    \mathcal{Y}_{\{2,3,4,5\}} = W_{5,\{4,5\}}\oplus W_{6,\{3,5\}}\oplus W_{4,\{3,4\}}\oplus W_{8,\{1,5\}}\oplus\\ W_{9,\{2,4\}}\oplus W_{10,\{2,3\}}
		\end{multline*}
\end{exmp}
\section{Performance analysis}
In this section we compare the performance of the proposed scheme with the schemes available in the literature for multi-access coded caching.

\subsection{Comparison with the HKD Scheme\cite{HKD}}
In Fig. \ref{HKDcomp_RK1}, the proposed scheme and the HKD scheme are compared  in terms of per user rate $\frac{R}{K}$ with respect to $\frac{M}{N}$, for $\frac{M}{N} < \frac{1}{r}$, and for different values of $r$. The expression for $R$ of centralized equivalent of the HKD scheme is given in \eqref{RateHKD}. Note that $r$ should divide $C$ for the centralized HKD scheme to exist. The number of users $K = C$ in the HKD scheme. Fig. \ref{HKDcomp_RK2} depicts variation of per user rate $\frac{R}{K}$ with respect to $\frac{M}{N}$ for the case when $\frac{M}{N} \geq \frac{1}{r}$. From Fig. \ref{HKDcomp_RK1}, it is seen that the proposed scheme achieves lower per user rate than the HKD scheme when $\frac{M}{N} < \frac{1}{r}$. For the case, $\frac{M}{N} \geq \frac{1}{r}$, in Fig. \ref{HKDcomp_RK2} it is is seen that the rate of the HKD scheme becomes $0$, while the per user rate of proposed scheme is slightly higher. However this is expected due to the cyclic wraparound setup considered in the HKD scheme and the placement in the caches such that the user gets all the $N$ files when $\frac{M}{N}\geq\frac{1}{r}$.

For the centralized equivalent of the HKD scheme, $F$ is given by $r{{\frac{C}{r}} \choose {t}}.$ 

In Fig. \ref{HKDcomp_rconst}, the variation of subpacketization $F$, with respect to $K$ is studied, keeping access degree $r$ constant for both schemes and the comparison is made for different values of $t$.
In Fig. \ref{HKDcomp_tconst}, the variation of subpacketization $F$, with respect to $K$ is studied, keeping $t = \frac{CM}{N}$ constant for both schemes and the comparison is made for different values of $r$.
It is seen that the subpacketization levels of proposed scheme is significantly lower than that of the HKD scheme for large number of users.

\begin{figure}[H]
\begin{center}
		\includegraphics[width=7.5cm,height=7cm]{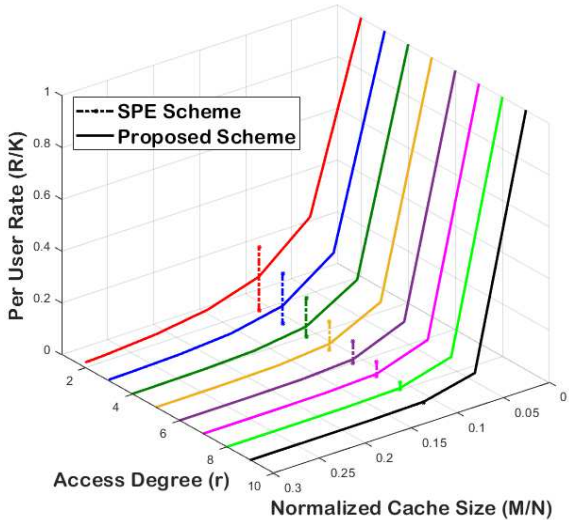}
		\caption {Comparison with the SPE scheme for number of caches, $C = 18$   for different values of $r$ i.e $r\in\{2,3,4,5,6,7,8\}$.}
		\label{SPEcomp_RK}
\end{center}
\end{figure}

\begin{figure}
\begin{center}
		\includegraphics[width=7.5cm,height=7cm]{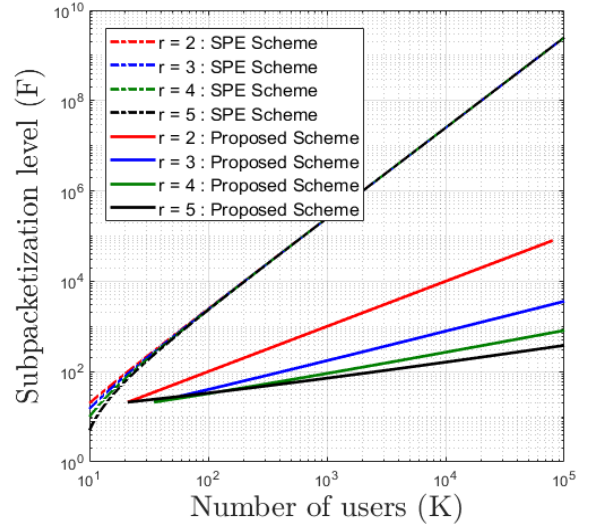}
		\caption {Comparison with the SPE scheme for different values of $r$ i.e $r\in\{2,3,4,5\}$ for the fraction of each file stored at each cache the same in both the schemes i.e. $\frac{M}{N} = \frac{2}{C}$.}
		\label{SPEcomp_MbyNconst}
\end{center}
\end{figure}

\begin{figure}[H]
	\centering
	\subfloat[$\frac{M}{N}  < \frac{1}{r}$]{
	\includegraphics[clip,width=0.8\columnwidth]{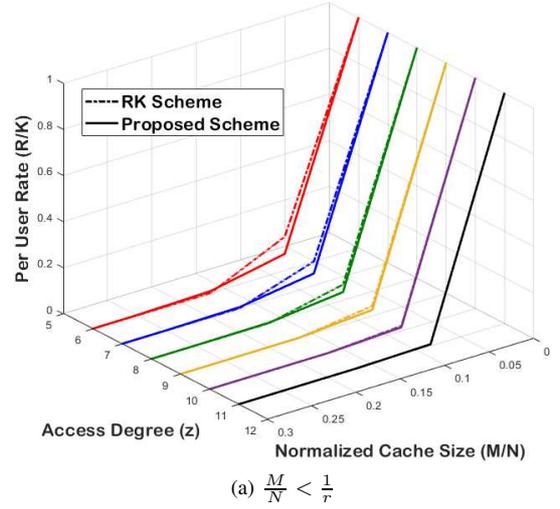}
	\label{RKcomp_RK1}}\\
	\subfloat[$\frac{M}{N}  \geq \frac{1}{r}$]{
	\includegraphics[clip,width=0.8\columnwidth]{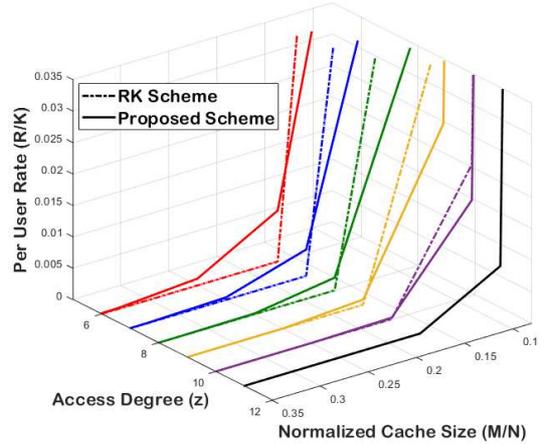}
	\label{RKcomp_RK2}}
\caption {Comparison with the RK scheme for the number of caches, $C = 12$ for different values of $r$ i.e $r\in\{6,7,8,9,10,11\}$.}
\label{RKcomp_RK}
\end{figure}

\subsection{Comparison with the SPE Scheme \cite{SBP}}

In Fig. \ref{SPEcomp_RK}, the proposed scheme and the SPE scheme for $\frac{KM}{N} = 2$ are studied in terms of per user rate $\frac{R}{K}$ with respect to $\frac{M}{N}$, by keeping access degree, $r=2$ for both the schemes. The expression for rate of the SPE scheme for $\frac{KM}{N} = 2$ is given in Theorem 1 of \cite{SBP}. For per user rate, it is normalized by number of users (For the SPE scheme $K = C$).
Since this scheme exists only for $\frac{KM}{N} = 2$, only these specific points are plotted for comparison. From Fig. \ref{SPEcomp_RK}, it is seen that proposed scheme supports lower per user rates. \\
The next comparison is in terms of subpacketization. In Fig. \ref{SPEcomp_MbyNconst}, variation of $F$ for the two schemes with respect to number of users $K$ is studied, for different values of $r$. It is seen that apart from supporting large number of users, lower subpacketization levels are obtained from the proposed scheme when compared to the SPE scheme.

\begin{figure}[H]
	\centering
	\subfloat[In the plot shown above, the two schemes are compared for different values of $r$ i.e $r\in\{2,3,4,5\}$, $t = \frac{CM}{N} = 2$.]{
	\includegraphics[clip,width=0.8\columnwidth]{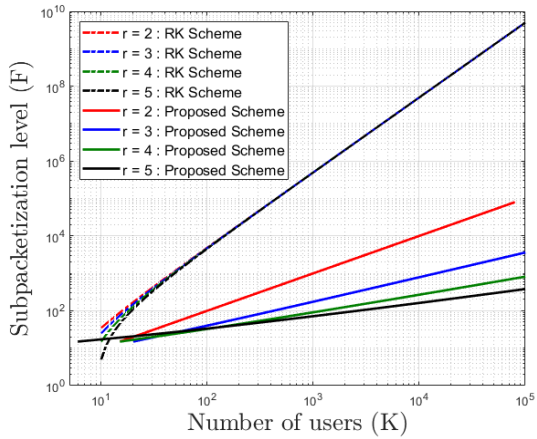}
	\label{RKcomp_MbyNconst}}\\
	\subfloat[In the plot shown above, the two schemes are compared for different values of $t = \frac{CM}{N}$ i.e $t\in\{2,3,4,5\}$, $r = 2$.]{
	\includegraphics[clip,width=0.8\columnwidth]{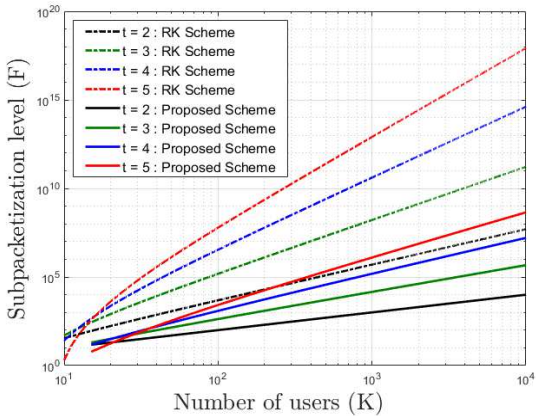}
	\label{RKcomp_zconst}}
\caption {Comparison with the  RK scheme in terms of subpacketization level ($F$).}
\end{figure}
\begin{table}[]
	\caption{Certain parameters for which SPE scheme performs better than proposed scheme}
	\begin{center}
		\renewcommand{\arraystretch}{2}
		\begin{tabular}{|c|c|c|c|}
			\hline
			\textbf{C} &\textbf{r}  &  \textbf{t} &  \textbf{$\frac{{\big(\frac{R}{K}\big)}_{Prop}}{\big({\frac{R}{K}\big)}_{SPE}}$}\\\hline\hline
			5	&	2	&	2	&	1.25	\\	\hline
		7	&	3	&	2	&	1.4	\\	\hline
		7	&	2	&	3	&	1.4	\\	\hline
		9	&	4	&	2	&	1.5	\\	\hline
		9	&	2	&	4	&	1.5	\\	\hline
		10	&	3	&	3	&	1.458	\\	\hline
		11	&	5	&	2	&	1.571	\\	\hline
		11	&	2	&	5	&	1.571	\\	\hline
		13	&	6	&	2	&	1.625	\\	\hline
		13	&	4	&	3	&	1.418	\\	\hline
		13	&	3	&	4	&	1.418	\\	\hline
		13	&	2	&	6	&	1.625	\\	\hline
		15	&	7	&	2	&	1.667	\\	\hline
		15	&	2	&	7	&	1.667	\\	\hline
		16	&	5	&	3	&	1.347	\\	\hline
		16	&	3	&	5	&	1.347	\\	\hline
		
		\end{tabular}
		\quad 
		\begin{tabular}{|c|c|c|c|}
			\hline
			\textbf{C} &\textbf{r}  &  \textbf{t} &  \textbf{$\frac{{\big(\frac{R}{K}\big)}_{Prop}}{\big({\frac{R}{K}\big)}_{SPE}}$}\\\hline\hline
			
			17	&	8	&	2	&	1.7	\\	\hline
			17	&	4	&	4	&	1.24	\\	\hline
			17	&	2	&	8	&	1.7	\\	\hline
			19	&	9	&	2	&	1.727	\\	\hline
			19	&	6	&	3	&	1.268	\\	\hline
			19	&	3	&	6	&	1.268	\\	\hline
			19	&	2	&	9	&	1.727	\\	\hline
			21	&	10	&	2	&	1.75	\\	\hline
			21	&	5	&	4	&	1.064	\\	\hline
			21	&	4	&	5	&	1.064	\\	\hline
			21	&	2	&	10	&	1.75	\\	\hline
			22	&	7	&	3	&	1.192	\\	\hline
			22	&	3	&	7	&	1.192	\\	\hline
			23	&	11	&	2	&	1.769	\\	\hline
			23	&	2	&	11	&	1.769	\\	\hline
			25	&	12	&	2	&	1.786	\\	\hline
			
		\end{tabular}
%
%
	\end{center}
	\label{SPEbetter}
\end{table}
For the case $r = \frac{C-1}{t}$, the authors in \cite{SBP}, propose an alternate scheme that achieves optimality, in the sense that the rate here exactly matches that of MAN scheme with each user having access to memory $rM$ each. Through search, we see that certain parameters corresponding to this case, can result in our scheme performing bad in terms of per user rate and subpacketization when compared to SPE scheme. Some of these parameters are depicted in Table \ref{SPEbetter}. In \cite{SBP}, for the scheme corresponding to  $r = \frac{C-1}{t}$,  $F = C$, while our subpacketization is $F_{SPE} = \binom{C}{t}$. When $r > t$, then proposed scheme is better in terms of subpacketization with respect to number of users. When $r = t$, both have same performance and when $r < t$, then SPE scheme is better in terms of subpacketization with respect to number of users. For the case when $r = C - 1$, $t = 1$, our setup matches cyclic setup and $r = \frac{C-1}{t}$. In this case the per user rate and subpacketization are same for both the schemes. For the case when $r = 1$, $t = C-1$, our setup matches the Maddah Ali Niesen scheme \cite{MaN} for dedicated caches and in this case also, the per user rate and subpacketization are exactly the same for both schemes.
It is interesting to observe that there does exist some parameters with $C- rt = 1$, where proposed scheme performs better than SPE scheme in terms of per user rate. Some of these parameters are given in Table \ref{SPEworse}.
\begin{table}[]
	\caption{Certain parameters for which proposed scheme is better compared to SPE scheme}
	\begin{center}
		\renewcommand{\arraystretch}{2}
		\begin{tabular}{|c|c|c|c|}
			\hline
			\textbf{C} &\textbf{r}  &  \textbf{t} &  \textbf{$\frac{{\big(\frac{R}{K}\big)}_{SPE}}{\big({\frac{R}{K}\big)}_{Prop}}$}\\\hline\hline
			25	&	6	&	4	&	1.097	\\	\hline
			25	&	4	&	6	&	1.097	\\	\hline
			26	&	5	&	5	&	1.205	\\	\hline
			29	&	7	&	4	&	1.274	\\	\hline
			29	&	4	&	7	&	1.274	\\	\hline
			31	&	10	&	3	&	1.006	\\	\hline
			31	&	6	&	5	&	1.537	\\	\hline
			31	&	5	&	6	&	1.537	\\	\hline
			31	&	3	&	10	&	1.006	\\	\hline
			33	&	8	&	4	&	1.47	\\	\hline
			33	&	4	&	8	&	1.47	\\	\hline
			34	&	11	&	3	&	1.064	\\	\hline
			34	&	3	&	11	&	1.064	\\	\hline
			36	&	7	&	5	&	1.94	\\	\hline
			36	&	5	&	7	&	1.94	\\	\hline
			37	&	12	&	3	&	1.123	\\	\hline
			37	&	9	&	4	&	1.685	\\	\hline
			37	&	6	&	6	&	2.131	\\	\hline
			37	&	4	&	9	&	1.685	\\	\hline
			37	&	3	&	12	&	1.123	\\	\hline
			40	&	13	&	3	&	1.182	\\	\hline	
		\end{tabular}
\end{center}
\label{SPEworse}
\end{table}
\subsection{Comparison with the RK Scheme\cite{RaK3}}

In this subsection, the proposed scheme is compared with the RK scheme scheme by varying $\frac{M}{N}$ and noticing its effect on per user rate. For the RK scheme the normalized lower bound of rate $\frac{R_{lb}}{K}$ is plotted. The expression for $R_{lb}$ is taken from Theorem 3 in \cite{RaK3}. Since this lower bound is valid for $r\geq \frac{C}{2}$, the comparison in Fig. \ref{RKcomp_RK} holds only for $r\geq \frac{C}{2}$. Also $K = C$ in the RK scheme. The schemes are compared for different values of $r$.
The per user rate for the proposed scheme is found to be better than that of the RK scheme for the range  $\frac{M}{N} < \frac{1}{r}$ from Fig. \ref{RKcomp_RK1}. 

In Fig. \ref{RKcomp_MbyNconst}, the two schemes are compared keeping M/N constant i.e. $\frac{M}{N} = \frac{2}{C}$. The two schemes have been compared for different values of $r$.
In Fig. \ref{RKcomp_zconst}, the two schemes are compared by keeping the access degree same in both the schemes i.e. $r = 2$. The two schemes have been compared for different values of $t$.
From these plots it can be concluded that significantly low subpacketization levels can be attained with the proposed scheme for large number of users compared with the RK scheme.


\subsection{Comparison with the CRD Scheme \cite{KNS}}

\begin{figure}
	\begin{center}
		\includegraphics[width=7.5cm,height=7cm]{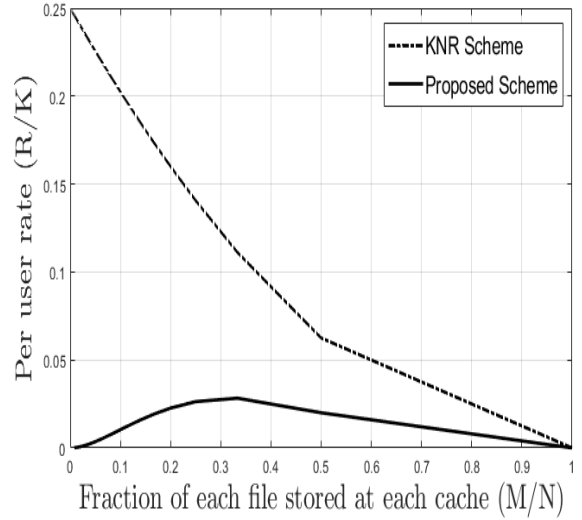}
		\caption {Comparison with the CRD scheme derived from affine planes.}
		\label{KNRcomp_RK}
	\end{center}
\end{figure}

The scheme from \cite{KNS} derived from the cross resolvable designs from affine planes is compared with proposed scheme in this section. 
The rate per user for the CRD scheme from affine planes is
$\frac{R}{K} = \frac{(n-1)^2}{4n^2}$ where $n$ is a prime or prime power. 
It can be seen from Fig. \ref{KNRcomp_RK}, that the rate per user supported, in the proposed scheme is significantly less compared to the scheme derived from CRD. 

\begin{figure}
	\centering
	\subfloat[Variation of Subpacketization with respect to n.]{
	\includegraphics[clip,width=0.8\columnwidth]{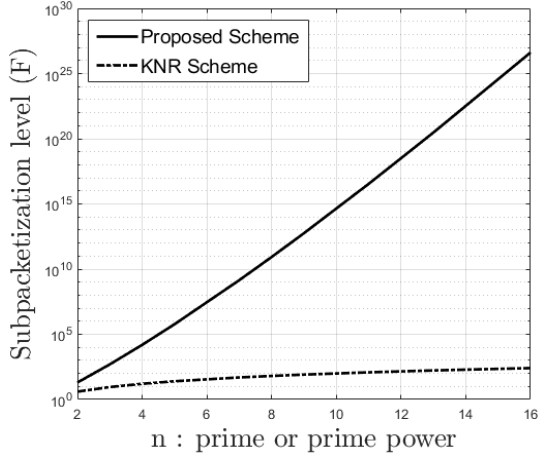}
	\label{KNRcomp_Fn}}\\
	\subfloat[Variation of Subpacketization with respect to number of users K.]{
	\includegraphics[clip,width=0.8\columnwidth]{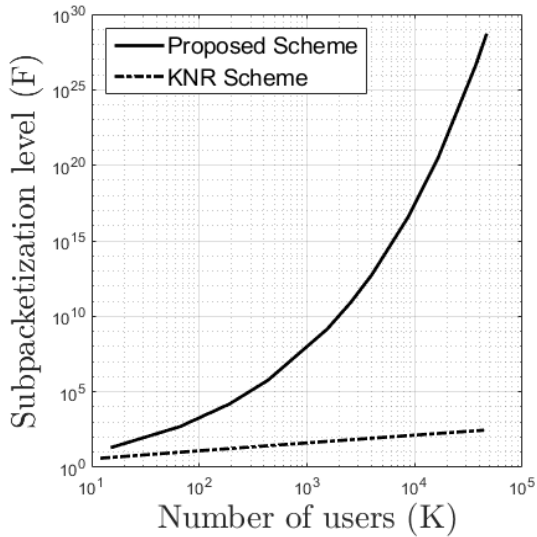}
	\label{KNRcomp_FK}}
\caption {Comparison with the  CRD scheme in terms of Subpacketization.}
\end{figure}

From \cite{KNS}, $\frac{M}{N} = \frac{1}{n}$, $C = n(n+1)$,$F = n^2$ and $K = \frac{n^3(n+1)}{2}$ for the multi-access scheme from CRDs derived from affine planes, where $n$ is a prime or prime power. 
In Fig. \ref{KNRcomp_Fn}, the number of caches $C$ and memory fraction $\frac{M}{N}$ is kept same for both schemes, and the two schemes are compared in terms of subpacketization levels $F$.
Since $C$ and $\frac{M}{N}$ are dependent only on the parameter $n$, for the CRD scheme from affine planes, the subpacketization values are plotted with respect to $n$.
In Fig. \ref{KNRcomp_FK}, the subpacketization levels are plotted with respect to number of users $K$.
From Fig. \ref{KNRcomp_Fn} and Fig. \ref{KNRcomp_FK}, it can be concluded that the proposed scheme does not perform well in terms of subpacketization levels when compared with the CRD scheme. But this is explained by the fact that the CRD scheme loses out in rate and gains in terms of subpacketization. Also, the number of users supported in the proposed scheme is more than that of the CRD scheme.


\subsection{Comparison with the  Cheng-Liang-Wan-Zhang-Caire (CLWZC) Scheme \cite{CLWZC}}

\begin{figure}
	\centering
	\subfloat[$\frac{M}{N}  < \frac{1}{r}$]{
	\includegraphics[clip,width=0.9\columnwidth, height = 6.5cm]{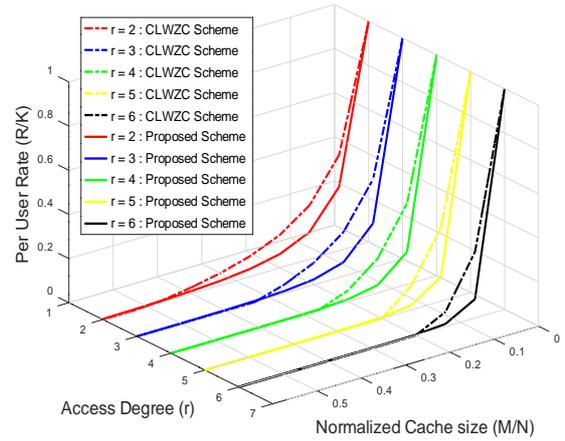}
	\label{CLWZC_RK1}}\\
	\subfloat[$\frac{M}{N}  \geq \frac{1}{r}$]{
	\includegraphics[clip,width=0.9\columnwidth, height = 6.5cm]{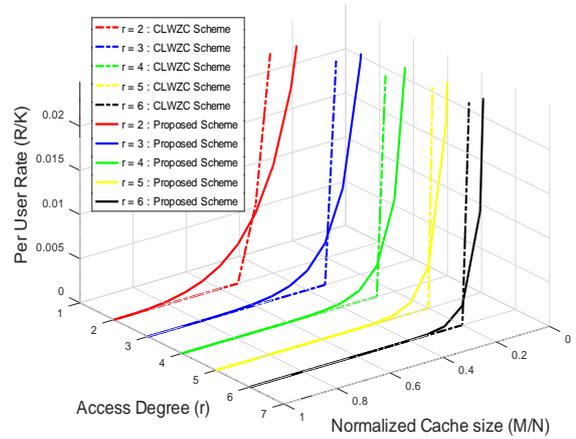}
	\label{CLWZC_RK2}}
\caption {Comparison with the CLWZC scheme for number of caches, $C = 15$ for different values of $r$ i.e $r\in\{2,3,4,5,6\}$.}
\end{figure}

In this subsection, the proposed scheme is compared with the CLWZC \cite{CLWZC} scheme scheme by varying $\frac{M}{N}$ and noticing its effect on per user rate. The expression of rate $R$ for \cite{CLWZC} scheme is taken from Theorem 1 in \cite{CLWZC}. The schemes are compared for different values of $r$.
The per user rate for the proposed scheme is found to be better than that of the CLWZC scheme for the range  $\frac{M}{N} < \frac{1}{r}$ from Fig. \ref{CLWZC_RK1}.

\begin{figure}
	\centering
	\subfloat[In the plot shown above, the two schemes are compared for different values of $t = \frac{CM}{N}$ i.e $t\in\{2,3,4,5\}$, $r = 2$.]{
	\includegraphics[clip,width=.85\columnwidth, height = 6.5cm]{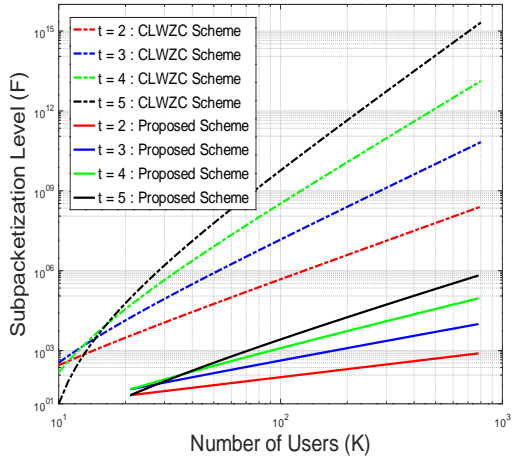}
	\label{clwzc_zconst}}\\
	\subfloat[In the plot shown above, the two schemes are compared for different values of $r$ i.e $r\in\{2,3,4,5\}$, $t = \frac{CM}{N} = 2$.]{
	\includegraphics[clip,width=0.85\columnwidth, height = 6.5cm]{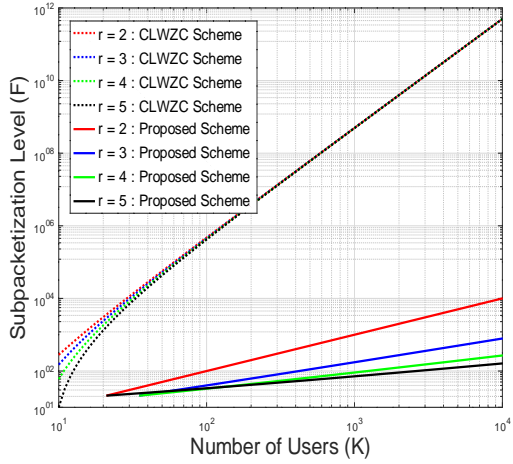}
	\label{clwzc_tconst}}
\caption {Comparison with the CLWZC scheme in terms of subpacketization ($F$).}
\end{figure}

In Fig. \ref{clwzc_tconst}, the two schemes are compared keeping $\frac{M}{N}$ constant i.e. $\frac{M}{N} = \frac{2}{C}$. The two schemes have been compared for different values of $r$.
In Fig. \ref{clwzc_zconst}, the two schemes are compared by keeping the access degree same in both the schemes i.e. $r = 2$. The two schemes have been compared for different values of $t$.
From these plots it can be seen that significantly low subpacketization levels can be attained with the proposed scheme when compared with the CLWZC scheme.

In Table \ref{tab1} the proposed scheme is compared with the CLWZC scheme for $t = \frac{CM}{N} = 1$. In can be observed that for $t=1$ and the access degree $r = C-2$ the proposed scheme achieves lower subpacketization level $(F)$ and supports larger number of users $(K)$ than the CLWZC scheme for the same fraction of each file at each cache $\frac{M}{N}$, fraction of each file each user has access to $\frac{M,}{N}$ and rate $(R)$.

\begin{table}
\caption{Comparison between CLWZC and Proposed scheme for $t = \frac{CM}{N} = 1$.}
  \begin{center}
  \renewcommand{\arraystretch}{2.5}
    \begin{tabular}{|c|c|c|}
    \hline
      \textbf{Parameters} &\textbf{CLWZC Scheme}  &  \textbf{Proposed Scheme}\\\hline\hline
      Number of Caches $(C)$ & $C$ &  $C$\\\hline
      \makecell{Number of Caches a \\ user has access to $(r)$} & $C-2$ &  $C-2$\\\hline
      Number of Users $(K)$ & $C$   &$\frac{(C)(C-1)}{2}$\\\hline
      \makecell{Fraction of each file\\ at each cache $\left(\frac{M}{N}\right)$} & $\frac{1}{C}$ & $\frac{1}{C}$ \\\hline
      \makecell{Fraction of each file\\ each user has access to } & $\frac{C-2}{C}$ & $\frac{C-2}{C}$ \\\hline
      Subpacketization level & $C\binom{C-t(r-1)}{t} = 3C$ & $\binom{C}{t} = C$\\[.2cm]\hline
      Rate (R) & $\frac{C-tr}{t+1} = 1$ & $\frac{\binom{C}{t+r}}{\binom{C}{t}}$ = 1 \\[.2cm] \hline 
    \end{tabular}
  \end{center}
\label{tab1}
\end{table}
\subsection{Comparison with the SR Scheme 1\cite{SRIC}}
\begin{figure}[]
	\centering
	\subfloat[$\frac{M}{N}  < \frac{1}{r}$]{
		\includegraphics[clip,width=0.8\columnwidth]{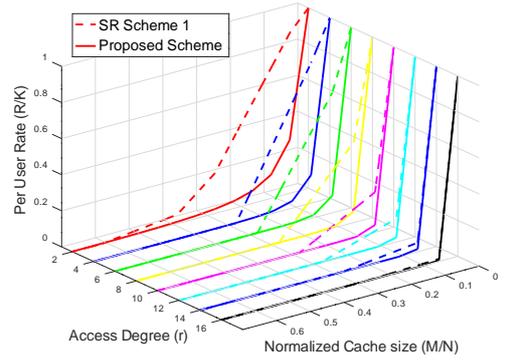}
		\label{}}\\
	\subfloat[$\frac{M}{N}  \geq \frac{1}{r}$]{
		\includegraphics[clip,width=0.8\columnwidth]{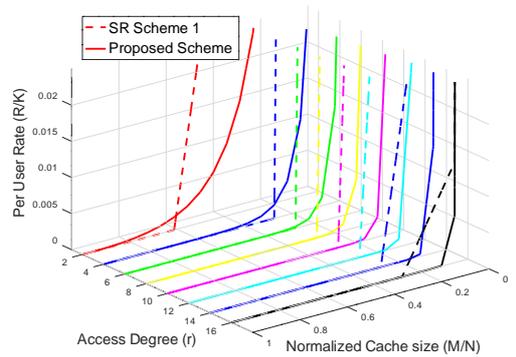}
		\label{sr1rk}}
	\caption {Comparison with the SR scheme 1 for the number of caches, $C = 18$ for different values of $r$ i.e $r\in\{2,4,6,8,10,12,14,16\}$.}
	\label{sr1rk2}
\end{figure}

Here, we compare our scheme to the SR Scheme 1.
The per user rate for the proposed scheme is found to be better than that of the SR scheme 1 scheme for the range  $\frac{M}{N} < \frac{1}{r}$ from Fig. \ref{sr1rk2}.

The subpacketization of the SR scheme 1 is atmost $F_{SPE} = C^2 = K^2$ as $C = K$, while ours is $\binom{C}{t}$ which can be less than, equal to, or greater than the number of users in the setup $\binom{C}{r}$.
So atleast for some choices of $t$ and $r$ the proposed scheme is better than the SR scheme 1 in terms of subpacketization. When $r = t$ for instance, the proposed scheme is better in terms of subpacketization with respect to number of users.

Note that in this subsection, we have not considered the case $r = \frac{C-1}{t}$, since it is already covered under comparison with SPE scheme.
\subsection{Comparison with the SR Scheme 2 \cite{SR}}
\begin{figure}[]
	\centering
	\subfloat[$\frac{M}{N}  < \frac{1}{r}$]{
		\includegraphics[clip,width=0.8\columnwidth]{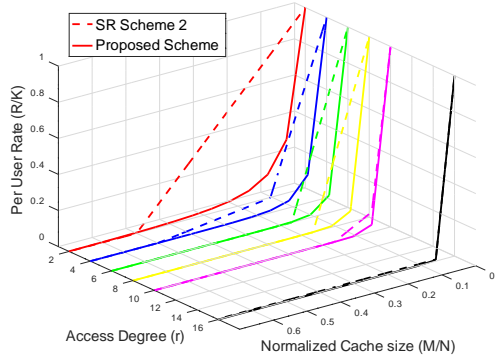}
		\label{}}\\
	\subfloat[$\frac{M}{N}  \geq \frac{1}{r}$]{
		\includegraphics[clip,width=0.8\columnwidth]{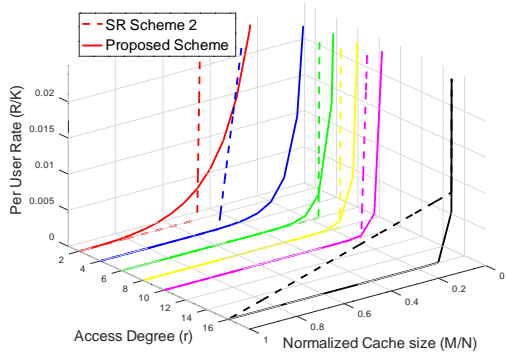}
		\label{sr2rk}}
	\caption {Comparison with the SR scheme 2 for the number of caches, $C = 18$ for different values of $r$ i.e $r\in\{2,4,6,8,10,16\}$.}
	\label{sr2rk2}
\end{figure}

Here, we compare our scheme to the SR Scheme 2.
The per user rate for the proposed scheme is found to be better than that of the SR scheme 2 scheme for the range  $\frac{M}{N} < \frac{1}{r}$ from Fig. \ref{sr2rk2}.
The subpacketization of the SR scheme 2 is $F_{SPE} = C = K$, while ours is $\binom{C}{t}$ which can be greater than, less than or equal to $\binom{C}{r}$, which says that our scheme can be better in terms of subpacketization than the SR scheme 2 for some choices of $r$ and $t$.
When $r = t$, both schemes have the same performance in terms of subpacketization with respect to number of users.  
\section{Conclusion}
We conclude that the proposed scheme is better than most existing schemes in literature since the number of users ($K$) supported is very large at relatively low subpacketization ($F$) levels. When $\frac{M}{N} < \frac{1}{r}$ and $r \neq \frac{C-1}{C{\frac{M}{N}}}$, the rate per user achieved is lesser than that of existing schemes. When $\frac{M}{N} < \frac{1}{r}$ and $r = \frac{C-1}{C{\frac{M}{N}}}$, the performance with respect to per user rate may be better or worse compared to existing schemes depending on the values of $C$, $r$ and $C\frac{M}{N}$. The proposed scheme exists for any multi-access network with $N$ files, $C$ caches equipped with memories of size $M$ file units each and access degree $r$. The scheme can be designed for any integer $\frac{CM}{N}$, and the rate points in between can be achieved through memory sharing, allowing convenience in designing.

\section*{Acknowledgment}
This work was supported partly by the Science and Engineering Research Board (SERB) of Department of Science and Technology (DST), Government of India, through J.C. Bose National Fellowship to B. Sundar Rajan.


\end{document}